\documentclass[final,twocolumn,merge,sort&compress]{elsarticle}

\usepackage{amssymb}
\usepackage[intlimits]{amsmath}
\usepackage{amsfonts}
\usepackage{dsfont}
\usepackage{subfigure}
\usepackage[usenames,dvipsnames]{color}
\usepackage{graphicx}
\usepackage{natbib}
\usepackage{multirow}
\usepackage{bbm}

\usepackage{slashed}
\usepackage{nicefrac}

\newcommand{\cO}{\mathcal{O}}
\newcommand{\cR}{\mathcal{R}}

\newcommand{\cT}{\mathcal{T}}

\newcommand{\be}{\begin{equation}}
\newcommand{\ee}{\end{equation}}
\newcommand{\bea}{\begin{eqnarray}}
\newcommand{\eea}{\end{eqnarray}}
\newcommand{\bi}{\begin{itemize}}
\newcommand{\ei}{\end{itemize}}

\begin{document}
\title{Asymptotically Safe $f(R)$-Gravity Coupled to Matter II: Global Solutions}

\author {Nat\'alia Alkofer}

\address{Institute for Mathematics, 
Astrophysics, and Particle Physics (IMAPP), Radboud University Nijmegen, \\
Heyendaalseweg 135, 6525 AJ Nijmegen, The Netherlands}

\date{\today}


\begin{abstract}
Ultraviolet fixed point functions of the functional renormalisation group equation 
for $f(R)$-gravity coupled to matter fields are discussed. The metric
is split via the exponential parameterisation into a background and a fluctuating metric, 
the former is chosen to be the one of a four-sphere.
Also when scalar, fermion and vector fields are included global quadratic solutions 
exist as in the pure gravity case for discrete sets of values for some endomorphism
parameters defining the coarse-graining scheme.
The asymptotic, large-curvature behaviour of the fixed point functions is analysed 
for generic values of these parameters. Examples for global numerical solutions 
are provided. A special focus is given to the question whether matter fields 
might destabilise the ultraviolet fixed point function. Similar to a previous 
analysis of a  polynomial,  small-curvature approximation to the fixed point 
functions different classes for such functions are found.
\end{abstract}

\maketitle
\section{Introduction}
\label{sec:intro}

Searching for a viable theory of quantum gravity is one of the most important open 
problems in theoretical physics. Many different approaches try to elucidate it
from various perspectives. In this letter, the asymptotic safety scenario for quantum 
gravity~\cite{Reuter:1996cp} will be employed based on a specific generalisation 
of Einstein's General Relativity:
an investigation  of $f(R)$-gravity minimally coupled to an 
arbitrary number of scalar, Dirac, and vector fields is 
discussed with a special focus on the study of global fixed functions,
the generalisation of non-Gau\ss ian fixed point (NGFP). Recently,
the flow equation used herein has been derived within the functional 
renormalisation group (FRG) for the effective average action, and
this equation has been solved to obtain the respective NGFP function in a 
polynomial,  small-curvature approximation \cite{Alkofer:2018fxj}, see also 
ref.\ \cite{PhDthesis}. These solutions provide the foundation of the 
here reported study. Results for the NGFP function for the pure gravity case 
within the employed version of the flow equation have been given recently in 
refs.\ \cite{Ohta:2015efa,Ohta:2015fcu}. Note that such solutions of NGFP functions
but for different truncations of the flow equation have been obtained in 
refs.~\cite{Reuter:2008qx,Benedetti:2012dx,Demmel:2012ub,Dietz:2012ic,Bridle:2013sra,
Dietz:2013sba,Demmel:2014sga,Demmel:2014hla,Demmel:2015oqa,Labus:2016lkh,
Dietz:2016gzg,Knorr:2017mhu,Falls:2017lst,Christiansen:2017bsy,deBrito:2018jxt}.
Hereby the characteristics of the solutions for these functions
differ significantly depending on the technical aspects of the respective work. 
Given the fixed functions' importance for the  
asymptotic safety scenario this requires further understanding. In the 
following it will be studied whether coupling matter might give an important
hint to resolve ambiguities.

Coupling matter to gravity within the asymptotic safety scenario has 
a long history, see refs.~\cite{Dou:1997fg,Percacci:2002ie,Narain:2009fy,
Daum:2010bc,Folkerts:2011jz,Harst:2011zx,Eichhorn:2011pc,
  Eichhorn:2012va,Dona:2012am,Henz:2013oxa,Dona:2013qba,
  Percacci:2015wwa,Labus:2015ska,Oda:2015sma,Meibohm:2015twa,Dona:2015tnf,
  Meibohm:2016mkp,Eichhorn:2016esv,Henz:2016aoh,Eichhorn:2016vvy,
  Christiansen:2017gtg,Eichhorn:2017eht,Christiansen:2017qca,
  Eichhorn:2017ylw,Eichhorn:2017lry,Eichhorn:2017egq,Christiansen:2017cxa,
Christiansen:2017bsy}, however, with mixed results. 
In ref.\ \cite{Alkofer:2018fxj} comprehensible estimates have been
provided which gravity-matter systems may give rise to NGFPs suitable for 
rendering the theory asymptotically safe. In this reference  
the flow equation has been derived  within a seven-parameter
family of non-trivial endomorphisms in the regularisation procedure. 
Herein this freedom will be exploited to show the existence of global
quadratic solutions. In ref.\ \cite{Alkofer:2018fxj} it was also shown 
that for vanishing endomorphisms gravity coupled to the matter 
content of the standard model of particle physics (and also many 
beyond the standard model  extensions) exhibit a NGFP whose 
properties are strikingly similar to the case of pure gravity: 
there are two UV-relevant directions, and the position and critical 
exponents  converge rapidly when higher powers of the scalar curvature 
 beyond the quadratic ones are included. Building on this 
result numerical solutions will be obtained for a global fixed 
function for the pure gravity as well as
for the gravity-matter system with standard model matter content.
Hereby the discussion of the singular points of the flow equation
and the asymptotic behaviour of the solution
for large scalar curvatures turns
out to be the crucial element. 

Based on generic features of the employed flow equation one can  show for 
global solutions a property already visible at the level of the 
polynomial approximations, namely that addition of fermions, stabilise 
an existing NGFP if the coarse graining operator is chosen as the
Laplacian. As the Standard Model is dominated by fermions therefore
a NGFP function for $f(R)$-gravity coupled to the Standard Model matter
content exists in this case. 

This letter is organised as follows: in sect.\  \ref{sec:RGE}  
 the derivation of the flow equation is briefly reviewed. 
In sect.\ \ref{sec:Sol4} the properties of the flow equation in four 
dimensions are discussed. The existence of global 
quadratic solutions is shown by constructing two explicit examples.
Furthermore, the asymptotic behaviour is analysed,  
and numerical solutions for two selected cases are presented.
In sect.\ \ref{sec:Concl} the results are summarised and 
conclusions are provided.

\section{RG equation for gravity and matter in $f(R)$ truncation}
\label{sec:RGE}

To make this letter self-contained the derivation of the 
flow equation given in ref.\ \cite{Alkofer:2018fxj} is briefly reviewed. 
It is based on the form of the 
FRG equations given in refs.\ \cite{Wetterich:1992yh,Morris:1993qb}, 
adapted to the case of gravity \cite{Reuter:1996cp},
\be\label{FRGE}
\partial_t \Gamma_k = \frac 1 2 \, {\mathrm {STr}} \left[ \left(\Gamma_k^{(2)} +        \cR_k\right)^{-1} \partial_t \cR_k \right] \, .
\ee
Here, $k$ is the RG scale and
$t = \ln(k/k_0)$ the RG ``time'' with $k_0$ being an arbitrary reference scale. 
$\Gamma_k$ denotes the effective average action and 
$\Gamma_k^{(2)}$ its second variation with 
respect to the fluctuation fields.  $\cR_k$ is a regulator introduced such 
that solving the flow equation effectively integrates out quantum fluctuations 
at and around the scale $k$. 
 
Throughout this work for the gravitational part of the effective average action a 
$f(R)$ truncation on $d$-spheres as background
will be employed. To be explicit, the corresponding total effective action reads
\be\label{Gammaans}
\Gamma_k = \Gamma_k^{\rm grav} + \Gamma^{\rm matter}_k \, ,
\ee
where $\Gamma^{\rm grav}_k$ is the gravitational part of the effective 
average action and $\Gamma^{\rm matter}_k$ contains the matter fields. 
The gravitational part of the action is assumed to be given by
\be
\Gamma_k^{\rm grav} = \int d^dx \, \sqrt{g} \, f_k(R)  \,\, + \Gamma_k^{\rm gf} 
+ \Gamma_k^{\rm gh}  \,  ,
\label{GammaGrav}
\ee
where $f_k(R)$ is an arbitrary, scale-dependent function of the Ricci scalar $R$,
and the action is supplemented by suitable gauge fixing and ghost terms. This 
sector is taken to be identical to the one studied in ref.\ \cite{Ohta:2015fcu}.
The matter sector is assumed to consist of $N_S$ scalar fields, 
$N_D$ Dirac fermions, and $N_V$ Abelian gauge fields.
The latter ones are fixed to Feynman gauge, and thus on curved backgrounds
the related ghosts are included. Matter self-interactions as well as the 
RG scale dependence of the matter wave-function renormalisations are neglected.

In a next step one splits the metric into a background and a fluctuating
 (quantum) part. In the following the exponential split
\be
g_{\mu\nu}= \bar g_{\mu\rho} ( e^h )^\rho{}_\nu\ 
\label{ExpSplit}
\ee
is used thereby avoiding any signature change even for large fluctuating 
fields. A detailed derivation of the  gravitational part of the RG equation
can be found in refs.\ \cite{Ohta:2015fcu,Ohta:2015efa}, see also
refs.\ \cite{Alkofer:2018fxj,PhDthesis}. 

Some of the global solutions for fixed functions presented here can only be 
obtained if additional endomorphisms in the regulator functions appearing in the
RG equations are introduced. To this end, 
a set of parameters $\alpha_{S,D,V,T}^{G,M}$, with the
subscript labelling the type of field and the superscript the gravity or matter sector, 
are introduced: the regulators are chosen to 
depend on  $\Box_{S,D,V,T}^{G,M} :=\Delta - \alpha_{S,D,V,T}^{G,M} \bar R$
where $\Delta$ is the Laplace operator,  and $\bar R$ is the positive curvature 
scalar of the background sphere. The labelling of subscripts refer to: 
$S$ scalar, $D$ Dirac, $V$ transverse vector, and $T$ transverse
traceless symmetric tensor field. 

In ref.\ \cite{Alkofer:2018fxj} all the traces appearing in the
 RG equation have been done by explicitly summing over the  eigenvalues of the 
Laplacian on the sphere in the different spin channels, 
for the respective expressions for these eigenvalues see, 
{\it e.g.},  ref.~\cite{Rubin:1984tc}. 
Using a Litim-type regulator \cite{Litim:2000ci,Litim:2001up}
\bea\label{RLitim}
R_k(z) &=&  (k^2-z) \theta(k^2 - z) \, , \\ 
\partial_t R_k(z) &=& 2 k^2 \theta(k^2 - z) \, ,
\nonumber
\eea
these sums are all finite. Subsequently, an additional smoothing operation
(namely averaging over two sums performed on the upper and lower limit of the
resulting ``staircase'' function)
are employed as part of the regularisation, see ref.\ \cite{Alkofer:2018fxj} for more 
details. 

Two widely used choices for the coarse graining operators termed ``Type I'' and 
``Type II'' (see \cite{Codello:2008vh} for detailed definitions and a discussion
of this typology) are given by the 
following choice of endomorphism parameters,
\begin{subequations}\label{endomorphism}
        \begin{align}
\label{endtypeI}
        & \mbox{Type I:}  \quad ~ \alpha^G_T = \alpha^G_S = \alpha^G_V  = 
 \nonumber \\ & \qquad \qquad \,\,\,
 =\alpha^M_D = \alpha^M_{V_1} = \alpha^M_{V_2} = \alpha^M_S = 0 \, , \\ 
\nonumber \\
\label{endtypeII}
                & \mbox{Type II:} \quad
                \alpha^G_T = - \tfrac{2}{d(d-1)} \, , \;
                \alpha^G_S = \tfrac{1}{d-1} \, , \;
                \alpha^G_V = \tfrac{1}{d} \, , \;  \nonumber \\ & \qquad \qquad \,\,\,
                \alpha^M_D = - \tfrac{1}{4} \, , \;
                \alpha^M_{V_1} = - \tfrac{1}{d} \, , \;
                \alpha^M_{V_2} = \alpha^M_S = 0 \, .
        \end{align}
\end{subequations}

The RG equation takes the form of a partial 
differential equation for the scale-dependent function $f_k(\bar R)$. As usual it
is advantageous to
formulate it in dimensionless variables $r = \bar R/k^2$ and $\varphi(r)=f(\bar R) /k^d$.
This leads to a separation of the ``classical'' scale dependence of $f(\bar R)$ from the
quantum one which reads in four dimensions:
\bea
\partial _t \Gamma_k &=& 
\int d^4 x \,\, \sqrt{g} \,\,  \partial _t f(\bar R)  \\   
&=&  V_4 \,\, k^4 \,\, \Bigl( \partial_t \varphi(r)
+ 4 \varphi(r) -2 r \varphi '(r) \Bigr)  , \nonumber 
\eea 
where $V_4=384 \pi^2 /\bar R^2$
is the volume of the $4$-sphere.
The flow equation is then given by:
\bea\label{fullRG4}
\dot{\varphi} + 4 \varphi - 2 r \varphi^\prime &=& \cT^{\rm TT} + \cT^{\rm ghost} +
 \cT^{\rm sinv}  \\   
&+& \cT^{\rm scalar} + \cT^{\rm Dirac} + \cT^{\rm vector} \, ,
 \nonumber 
\eea
where
\begin{subequations}\label{gravflow}
        \begin{align}\label{gravTT}
        \cT^{\rm TT} = & \,
        \frac{5}{2 (4\pi)^2} \, \frac{1}{1 + \left(\alpha^G_T + \tfrac{1}{6}\right)r} 
\left(1 + \left(\alpha^G_T - \tfrac{1}{6}\right)r\right) \\
& \qquad  \qquad  \qquad \quad \times  \left(1 + \left(\alpha^G_T - \tfrac{1}{12}\right)r\right)
\nonumber  \\
 \nonumber 
                &+ \frac{5}{12 (4\pi)^2} \, \tfrac{\dot{\varphi}^\prime + 2 \varphi^\prime - 2 r \varphi^{\prime\prime}}
{\varphi^\prime} 
\left(1 + \left(\alpha^G_T - \tfrac{2}{3}\right)r\right) \\
& \qquad  \qquad  \qquad \qquad \quad  \times
\left(1 + \left(\alpha^G_T - \tfrac{1}{6}\right)r\right) \, ,
 \nonumber  \\ 
\label{gravsinv}
        \cT^{\rm sinv} = & \,
        \frac{1}{2 (4\pi)^2}
        \tfrac{ \varphi^{\prime\prime}}{\left(1+ \left(\alpha^G_S - \tfrac{1}{3}\right)r\right) 
\varphi^{\prime\prime} + \tfrac{1}{3} \varphi^\prime} \\
& \qquad \times 
        \left(1 + \left(\alpha^G_S - \tfrac{1}{2}\right)r\right)
        \left(1 + \left(\alpha^G_S + \tfrac{11}{12}\right)r\right)
       \nonumber   \\ 
\nonumber
         &+ \frac{1}{12 (4\pi)^2}
         \tfrac{\dot{\varphi}^{\prime\prime} - 2 r \varphi^{\prime\prime\prime}}
{\left(1+ \left(\alpha^G_S - \tfrac{1}{3}\right)r\right) \varphi^{\prime\prime} + \tfrac{1}{3} \varphi^\prime}
\\ \nonumber & \qquad \times 
\left(1 + \left(\alpha^G_S + \tfrac{3}{2}\right)r\right)
         \left(1 + \left(\alpha^G_s - \tfrac{1}{3}\right)r\right)\\
& \qquad \times 
         \left(1 + \left(\alpha^G_S - \tfrac{5}{6}\right)r\right) \, , \nonumber       \\
        \cT^{\rm ghost} = & \, - \frac{1}{48 (4\pi)^2} \, \frac{1}{1 + (\alpha^G_V - \tfrac{1}{4})r} \,
\label{gravghost} \\
& \qquad \times 
 \Bigl( 72 + 18 r (1 + 8 \alpha^G_V)   \nonumber    \\
& \qquad \qquad 
 - r^2 (19 - 18 \alpha^G_V - 72 (\alpha^G_V)^2) \Bigr) \, ,
\nonumber 
        \end{align}
\end{subequations}
and
\begin{subequations}\label{matterflow}
        \begin{align}
\cT^{\rm scalar} = & \, \frac{N_S}{2 (4\pi)^2} \, \frac{1}{1 + \alpha_S^M r} \, 
\left( 1 + \left( \alpha_S^M + \tfrac{1}{4} \right) r \right)  \\
& \qquad  \qquad  \qquad \,\,\, \times 
\left(1 + \left( \alpha_S^M +  \tfrac{1}{6} \right) r\right) \, , \nonumber    \\ 
\label{Tdirac}
\cT^{\rm Dirac} = & \, - \frac{2 N_D}{(4\pi)^2} \,
\left(1 + \left(\alpha^M_D + \tfrac{1}{6}\right)r\right) \, , \\
 \nonumber  
\cT^{\rm vector} = & \, \frac{N_V}{2 (4\pi)^2} \, \bigg(
\tfrac{3}{1 + \left( \alpha^M_{V_1} + \tfrac{1}{4} \right) r}
\left(1 + \left(\alpha^M_{V_1} + \tfrac{1}{6} \right) r \right)  \\
& \qquad  \qquad  \qquad \qquad \times 
\left(1 + \left(\alpha^M_{V_1} + \tfrac{1}{12} \right)r\right) \nonumber  \\ 
& \qquad \qquad 
- \tfrac{1}{1 + \alpha^M_{V_2} r} \left( 1 + (\alpha^M_{V_2} + \tfrac{1}{2}) r \right)
\nonumber  \\ 
& \qquad \qquad \qquad \qquad  \times
 \left(1 + ( \alpha^M_{V_2} - \tfrac{1}{12}) r\right)
\bigg) \, .   
        \end{align}
\end{subequations}

The first  line of eq.\ (\ref{fullRG4}) stems from the gravitational sector and
depends correspondingly on the endomorphism parameters $\alpha^G_{S,V,T}$. 
The second line originates 
from the matter part: the contributions from the transverse vector and 
scalar ghost fields are proportional to $N_V$. In addition there are the ones
of the Dirac and the scalar fields. 

The common factor $1/(4\pi)^{2}$ can be removed from the coefficients defined 
in eqs.\ (\ref{gravflow}) and (\ref{matterflow}) by a suitable rescaling of $\varphi(r)$, 
and this is assumed in the following ({\it cf., e.g.}, eq.\ \eqref{quadAnsatz} below).

\section{Flow equation and fixed functions in four dimensions}
\label{sec:Sol4}

\subsection{Discussion of the flow equation}

Due to the chosen regulator the coefficient $\cT^{\rm Dirac}$
shows a peculiarity, it is only linear in the curvature. 
During derivation it is also a ratio of a quadratic numerator and a 
linear denominator like the other
 coefficients, however, for the chosen regularisation
procedure (and only for this one amongst the ones used in ref.\ \cite{Alkofer:2018fxj}) 
this denominator cancels against one of the two numerator terms
 to yield
$\cT^{\rm Dirac}\propto 2 (1+(\alpha_D + \nicefrac 1{6})r)$. 

Although simpler than the other terms the coefficient for the Dirac fields 
already displays a qualitative difference when changing the related endomorphism parameter. 
The allowed interval for $\alpha_D^M$ is  $-1/4 \le \alpha_D^M \le 0$
with the lower end corresponding to a  type-II-- and the upper end to a type-I--regulator.
It is plain that therefore the sign of the linear term depends on this parameter, 
and this will qualitatively change how fermions contribute to the flow equation.
This property will be important when discussing the solutions for fixed functions.

Solving the non-linear partial differential equation \eqref{fullRG4} for flows of the
function $\varphi(r)$ is an extremely complicated task. The necessary 
first step in such an analysis
is calculating its fixed functions, the generalisation of fixed points.
Those are the solutions of the ordinary differential 
equation obtained by setting $\partial_t \varphi(r)=0$ and thus 
$\partial_t \varphi'(r)=0=\partial_t \varphi''(r)$ in eq.~(\ref{fullRG4}). 
To distinguish them from the general scale-dependent function $\varphi(r)$  
a fixed function will be denoted as usual by $\varphi^\star(r)$ in the following.

In its normal form $\varphi^{\star \, \prime\prime\prime}  = \ldots $ the flow equation
has the following singularities: first, from the term proportional to 
$\varphi^{\star \, \prime\prime\prime}$ in \eqref{gravsinv}
\bea 
\label{fixedsing}
r^{\rm sing}_1 &=&  - \frac{1}{\alpha^G_S + \tfrac{3}{2}} \, , \, \, \,
r^{\rm sing}_2 = 0 \, , \\ 
r^{\rm sing}_3 &=& - \frac{1}{\alpha^G_S -\tfrac{5}{6}} \, , \,\,\,
r^{\rm sing}_4 = - \frac{1}{\alpha^G_S - \tfrac{1}{3}}  \, ,
\nonumber
\eea
and, second, from the denominators in the expressions \eqref{gravflow} and 
\eqref{matterflow}. Hereby the extrema of $\varphi^{\star}(r)$ via the first term
in \eqref{gravTT} and the denominators in \eqref{gravsinv} are moving 
singularities. As seen below one can arrange the parameters and the solutions
such that the moving singularities are canceled against the numerators. 
Note that the singularity at vanishing curvature, $r^{\rm sing}_2 = 0$, 
reflects the non-smooth transition from a sphere to a flat space. 

In the pure gravitational sector global fixed functions which are polynomials of quadratic 
order have been found and described in ref.\ \cite{Ohta:2015fcu}. 
For them, the third derivative vanishes and thus
the second summand of the expression \eqref{gravsinv} does not contribute.
 In all other non-trivial solutions
this term (which stems from the conformal mode) determines 
the structure of the differential equation in the normal form because
in this and only this term a third-order derivative, {\it i.e.}, $\varphi'''(r)$, 
appears.

\subsection{Global solutions for fixed functions}
\label{sec:Sol}

\subsubsection{Global quadratic solutions}

As already mentioned above, global solutions of quadratic order,
\be
\varphi^\star (r) = \frac 1 {(4\pi)^2} \,\, (g_0^\star + g_1^\star r + g_2^\star r^2)
\label{quadAnsatz}
\ee
are special. Hereby, $g_1^\star$ needs to assume a negative value.
To understand why one requires this for a polynomial Ansatz
one writes the action such 
that the Einstein-Hilbert action in standard notation is contained, 
\be
f(R) = \frac {\Lambda_k}{8\pi G_k} - \frac R {16 \pi G_k} + {\cal O} (R^2) \, , 
\ee
which allows one to identify 
\be
\Lambda_k = - \frac {g_0}{2g_1} k^2 \qquad {\mathrm {and}} \qquad G_k= - \frac \pi {k^2 g_1} \, ,
\ee
respectively.
Thus, a positive value for Newton's constant requires a negative value of $g_1$. As the RG flow for 
Newton's constant cannot cross the zero, also its fixed point value must be positive for an 
acceptable solution, and thus $g_1^\star <0 $. 

The fact that the constant term in the polynomial expansion, 
$g_0$, does not 
appear on the right hand side of the flow equation (\ref{fullRG4}) allows for a simple 
estimate how it is changed by the presence of matter fields. This in turn permits to 
estimate the influence 
of matter onto the cosmological constant within the present setting:
$g_0^{complete} \approx g_0^{gravity} + \frac 1 4 N_V +\frac 1 8 N_S -  \frac 1 2  N_D$,
 for more details see \cite{Alkofer:2018fxj}. Furthermore, the related critical exponent
(which is as usual defined as the negative of the eigenvalue of the stability matrix of the
linearised flow equation)
is always $\Theta_0=4$.

Without matter fields, {\it i.e.}, for $N_S=N_D=N_V=0$, five different
solutions for a globally quadratic fixed function have been identified in ref.\ \cite{Ohta:2015fcu}. 
In case $\varphi^\star(r)$ is a polynomial the differential equation determining it can be written as
\be
\frac {{\cal P}_{num}(r) }{{\cal P}_{den}(r) } = 0 ,
\ee
{\it i.e.}, as the requirement that the ratio of two polynomials vanish. This can be solved 
in two steps: first, solve for ${{\cal P}_{num}(r) }=0 $, and second, keep only those
solutions where all roots of ${{\cal P}_{den}(r) }$ ({\it i.e.}, the potential singularities 
of this equation) coincide with roots of the numerator.

In the case of a quadratic Ansatz for the fixed function, $ {{\cal P}_{num}(r) }$ is a 
fifth-order polynomial,\footnote{\label{FootCanc} 
The l.h.s\ of the flow equation (\ref{fullRG4}) is for the Ansatz (\ref{quadAnsatz}) not 
a polynomial of order $N=2$ as na\"ively expected because the term proportional to $r^2$ cancels:
$4 \varphi(r) -2 r \varphi '(r) = 4g_0+2r g_1 $.}
 and its six coefficients can be determined by a discrete set of values 
for $g_0^\star$, $g_1^\star$, $g_2^\star$, $\alpha_T^G$, $\alpha_V^G$ and $\alpha_S^G$,
see ref.\ \cite{Ohta:2015fcu}. 
Quite surprisingly, in all five solutions found in this reference 
the potential singularities given by the zeros of the denominator are canceled by the numerator. 
On the other hand, for two of these five solutions the eigenperturbations lead to a differential 
equation with four instead of three fixed singularities, and therefore such eigenperturbations cannot 
exist globally. For another of these five solutions $\alpha_T = (11 + \sqrt{265})/54 \approx 0.505 > 2/3$,
{\it i.e.}, the inequality for a positive argument of the regulator function is violated. This leaves
us with two solutions, and the corresponding values for the parameters are given in the respective 
first lines of tables \ref{tabQ1} and \ref{tabQ5}. The exact values of these parameters are, 
respectively,
{\begin{align}
\label{Q1}
\alpha_S^G=\tfrac{5\sqrt{265}-73}{216} & , & \alpha_V^G= \tfrac{67-2 \sqrt{265}}{108} & , &
\nonumber \\
\alpha_T^G=\tfrac{11-\sqrt{265}}{54}  & , & 
g_0^\star=\tfrac{49+\sqrt{265}}{96} & , & \\
g_1^\star =- \tfrac{4141+121\sqrt{265}}{5184} & , &g_2^\star= \tfrac{67795 + 3583 \sqrt{265}}{279936},
\nonumber
\end{align}}
or
$$
\alpha_S^G=-\tfrac 3 {47} \, ,\,\,\, \alpha_V^G = - \tfrac {83}{564}  \, , \,\,\, 
\alpha_T^G=- \tfrac {53}{94}\, , \,\,\, 
$$
\be g_0^\star = \tfrac {89}{72} \, , \,\,\,
g_1^\star = - \tfrac {101}{94}  \, ,\,\,\, g_2^\star = \tfrac {1414}{6627} \,  . \label{Q5}
\ee

As this will be important below we also give the value for the minimum of the fixed functions
\be
r_{min} = - \frac {g_1^\star}{2g_2^\star} = 
\begin{cases}
\frac 3 {20} ( 25 - \sqrt{265}) \approx 1.3082 \, , &\\
\\
\frac{141}{56} \approx 2.5179 \, .& \\
\end{cases}
\ee
As a matter of fact, for a  global quadratic solution one can rewrite the equation
for  the fixed function such that the parameters $g_i^\star$ appear only in the ratio
$r_{min} = - {g_1^\star}/{2g_2^\star}$  on the left hand side
because then some of the expressions in eqs.\ \eqref{gravTT} and \eqref{gravsinv}
simplify:
\be
 \frac{ \varphi^\prime -  r \varphi^{\prime\prime}} {\varphi^\prime}  = 
 \frac{r_{min}}{r_{min}-r}
\ee
and
\be
\frac{ \varphi^{\prime\prime}}{\left(1+ \left(\alpha^G_S - \tfrac{1}{3}\right)r\right) 
\varphi^{\prime\prime} + \tfrac{1}{3} \varphi^\prime} = 
 \frac 1{1 + \alpha_S^G r - r_{min}/3} \, .
\ee

For the solution (\ref{Q1}) one of the zeros of second summand of \eqref{gravTT}
occurs exactly at $r_{min}$ and thus the 
potential singularity is canceled. The singularity in the scalar term occurs at negative values of 
$r$ and is thus of no concern. For the solution (\ref{Q5}) the potential pole due to the scalar term
appears also exactly at $r_{min}$, and the same is true for the first term in
\eqref{gravTT} and the term \eqref{gravghost}.
With these values of endomorphism parameters, for the pure gravity case, the four terms of the 
left hand side  conspire to yield 
\be
\frac 1 {(4\pi)^2} \left(\frac {89}{18} - \frac {101}{47} r \right) 
\ee
which, of course, solves then the equation for the fixed function for the parameters $g_0^\star$ and 
$g_1^\star$ given in eq.~(\ref{Q5}).

The usefulness of the above considerations becomes immediately clear when adding fermions, {\it i.e.},
when adding 
\be
\frac {-2N_D} {(4\pi)^2} \left(1+ (\alpha_D^M + \frac 1 { 6})  \,\, r \right)  \, .
\ee
A global quadratic solution can be now easily obtained by keeping the ratio 
$  {g_1^\star}/{g_2^\star}$ 
and thus $ r_{min} $ fixed.  One simply keeps the values of the endomorphism 
parameters in the gravity sector and substitutes
\bea
g_0^\star & \to &  g_0^\star - \frac {N_D}{2} \nonumber  \\
g_1^\star & \to &  g_1^\star - (\alpha_D^M + \frac 1 { 6}) N_D 
 \nonumber  \\
 &&=
g_1^\star + \begin{cases}
 \frac 1 {12} N_D & {\mathrm {type \, II\, reg.}} \\
- \frac 1 6 N_D  & {\mathrm {type \, I~\, reg.}} \\ 
\end{cases} \label{RulesForFermions}\\
g_2^\star & \to &  g_2^\star + \frac 1 {2r_{min}}  (\alpha_D^M
+ \frac 1 { 6}) N_D 
 \nonumber  \\
 &&=
g_2^\star - \frac 1 {2r_{min}} \begin{cases}
 \frac 1 {12} N_D  & {\mathrm {type \, II\, reg.}} \\
- \frac 1 6 N_D  & {\mathrm {type \, I~\, reg.}} \\ 
\end{cases} \nonumber
\eea
This proves to be always possible independent of  whether the coefficient of
the linear term is negative as, {\it e.g.}, for the type I regulator, or positive
as, {\it e.g.}, for the type II regulator. However, in the latter case  the value 
of $g_1^\star$ will change sign, and thus the solution becomes unphysical.
 
If one uses now the type-II regulator for the fermions there will be a critical value of $N_D$ where 
$g_1^\star$ becomes positive, for the solution (\ref{Q1}) this value is $N_D=14.1$ 
whereas for the solution
(\ref{Q5}) it is $N_D=12.9$. 
If these values are exceeded the minimum turns to a maximum (but stays at
the same location) and the values of $g_1^\star$ and $g_2^\star$ change sign. 
Therefore, if a type-II regulator is used for  fermions one can add only a finite number of 
them and keep a physically meaningful solution in agreement with the results obtained 
already in the polynomial approximation \cite{Alkofer:2018fxj}.

Adding now scalar and/or vector fields it turns out that one cannot fix 
the parameters $\alpha_S^M=\alpha_{V2}^M=0$ and $\alpha_{V1}^M=-1/4$, {\it i.e.}, 
to their respective type-II values.
Although then no new singularities arise in the matter sector one can easily convince 
oneself that one obtains then 
for the numerator polynomial the degree six, and thus seven equations for six variables
because the expressions $\cT^{\rm scalar}$ and$ \cT^{\rm vector} $ 
in \eqref{matterflow} are of quadratic order.  A similar
situation arises, namely eight equations for seven variables etc., if one fixes only 
one or two 
of the three parameters to the respective type-II value.  Basically the same remark applies for fixing to 
type-I values

Exploring the possibility of adjusting the parameters $\alpha_S^M$ and $\alpha_{V1,2}^M$
to keep a global quadratic solution
one notes first that adding fermions is always straightforward by applying the rule 
(\ref{RulesForFermions}). 
For finding the endomorphism parameters which lead to a quadratic solution 
it proves to be easier to add scalar then vector fields. 
To obtain a solution with the standard model field content the following strategy has been used: 
first, add 45/2 Dirac fields 
(according to standard model matter content) with
type-I regulator by applying (\ref{RulesForFermions}) to the solution
(\ref{Q1}) and verify this numerically.
Second, on the top of this four scalar fields are added and the corresponding parameter
$\alpha_S^M$ is determined. From there on one increases $N_V$ in small steps until 
the standard model value 12 was reached. The results for pure gravity, gravity plus 
fermions, gravity plus fermions and scalars as well as for gravity plus standard model
matter content are  displayed in tables
\ref{tabQ1} and \ref{tabQ5}. In all cases one obtains $\alpha_S^M=\alpha_{V2}^M$.

\begin{table*}[ht]
\scalebox{0.94}{
\begin{tabular}{||c||c|c|c|c|c|c||c|c|c|c||c||}
\hline
$(N_S,N_D,N_V)$&$\alpha_S^G$ &$\alpha_V^G$ &$\alpha_T$ &$\alpha_S^M$ & 
$\alpha_D^M$  &$\alpha_{V1}^M$ &
$g_0^\star$ & $g_1^\star$ & $g_2^\star$ & $r_{min}$  & $\Theta_0, \Theta_1$ \\
\hline
(0,0,0)      & .0389 & .3189 & -.0978 & -           & -    & -        & .6800 & -1.179  & .4505 & 1.308 & 4,  2.02 \\ 
(0,45/2,0) & .0389 & .3189 & -.0978 & -           &  0  & -         & -10.57 & -4.929  & 1.884  & 1.308 & 4,  1.98 \\
(4,45/2,0) & -.0819 & .0389 & -.3778 & -.2111 &  0  & -         & -9.970 & -5.078  & 1.382 &  1.837 & 4,  2.35 \\
(4,45/2,12)&-.0190 & .1603 & -.2563 & -.0897 &  0  & -.3397 & -6.702 & -8.630  & 1.825 & 2.364 & 4,  2.36 \\
\hline
\end{tabular}
}
\caption{ Quadratic solutions for the fixed function with different matter content derived from
the pure gravity solution (\ref{Q1}).
\label{tabQ1}}
\end{table*}
\begin{table*}[ht]
\scalebox{0.935}{
\begin{tabular}{||c||c|c|c|c|c|c||c|c|c|c||c||}
\hline
$(N_S,N_D,N_V)$&$\alpha_S^G$ &$\alpha_V^G$ &$\alpha_T$ &$\alpha_S^M$ &$\alpha_D$  &$\alpha_{V1}^M$ &
$g_0^\star$ & $g_1^\star$ & $g_2^\star$ & $r_{min}$  & $\Theta_0 , \Theta_1$ \\
\hline
(0,0,0)      & -.0638  & -.1472 & -.5638 & -          & -    & -         & 1.236  & -1.074 & 0.2134 & 2.518 & $\approx$ 16, 4 \\ 
(0,45/2,0) & -.0638  & -.1472 & -.5638 & -          &  0  & -          & -10.01 & -4.824 & 0.9581 & 2.518 &   4, 2.98 \\
(4,45/2,0) & -.0554  & -.1388 & -.5550 & -.3855 &  0  & -          & -9.415 & -4.637 & 0.9014 & 2.572 &   4, 3.05 \\  
(4,45/2,12)& -.0308 & -.1140 & -.5308 & -.3644 &  0  & -.6143 & -5.813 & -9.368 & 1.705   & 2.746 &   4, 2.8  \\
\hline
\end{tabular}
}
\caption{ Quadratic solution for the fixed function with different matter content derived from
the pure gravity solution (\ref{Q5}).
\label{tabQ5}}
\end{table*}

It has to be emphasised that a solution with a positive value for Newton's constant could be found
because the type-I value $\alpha_D^M=0$ was used for the fermionic term. 
The stabilising effect of the type-I regulated fermions is very much needed.
{\it E.g.}, the solution with no fermions at all but four scalars and twelve vectors which follows from the 
ones given in table \ref{tabQ1} possesses a negative value for Newton's constant. 
For the solution (\ref{Q5}) one obtains an interesting effect of the fermions for the critical exponents: 
the pure gravity solution has a large critical exponent 
which we estimate to be around 16. Adding now type-I regulated fermions brings this one down to three 
(which also restores the order such that the critical exponent 4 related to the cosmological constant is 
the largest one). Adding scalars on top of gravity and fermions slightly increases the 
values of the second critical exponent 
but has overall not much effect. The same can be said about the vector fields.
All other critical exponents $\Theta_{2,3,\ldots}$ are always negative, respectively,
possess a negative real part.

In summary, two solutions have been found with endomorphism parameters adjusted such that 
a global quadratic solution exist for matter up to the Standard Model matter content. This 
worked  because  a type-I regulator for the fermions has been used. At least for the 
type of solutions discussed here type-II regulated fermions quite efficiently lead to 
a change of sign of Newton's constant and 
thus outside the class of physically accepted solutions.

\goodbreak

\subsection{Asymptotic behaviour for large curvature}

Studying the asymptotic behaviour for large curvature $r$ serves within this investigation 
two purposes. On the one hand, this knowledge will be employed when numerically solving 
for a fixed function. On the other hand, it will allow to identify a destabilising influence 
of matter fields without actually searching for a numerical solution.

As shown below the possible  leading asymptotic behaviour for $r\gg 1$ is  
either $\propto r^2$ or $\propto r^2 \ln r$ ({\it cf.}, ref.\ \cite{Dietz:2012ic})
depending on the values of the endomorphism 
parameters.
The left hand side of eq.~\eqref{fullRG4} at the NGFP
reduces to a constant  plus linear term if $\varphi^\star (r)$ is a quadratic function due to a 
cancelation (see  footnote~\ref{FootCanc}), and it becomes a quadratic polynomial
if a term proportional to $r^2 \ln r$ is added. 
 
Quite obviously cancelations in differences between terms play a significant role. 
Therefore the most straightforward way to proceed is to infer the large curvature behaviour 
term by term. In this respect the simplest term is $\cT^{\rm Dirac}$ \eqref{Tdirac}. 
It is,  in the presence of a non-vanishing quadratic term 
on the left hand side of the flow equation, subleading because 
it is a linear function in $r$. 
As the scalar matter term $\cT^{\rm scalar}$ and the contribution from the gauge
ghost behave identical they can be discussed together. One clearly sees a qualitative
difference for $\alpha_S^M\not =0$, resp., $\alpha_{V2}^M\not= 0$ for which the 
asymptotic behaviour of the corresponding terms is linear, versus for vanishing 
parameter (which includes type-I and type-II coarse graining) for which the
asymptotic behaviour is quadratic. In the first case these two terms provide
singularities at $r=-1/\alpha_S^M$ and $r=-1/\alpha_{V2}^M$, respectively.
In the latter case one has, of course, no singularities.

As for the transverse vector matter fields one has to distinguish between the
type-II case $\alpha_{V1}^M =-1/4$ for which there is no singularity but a 
quadratic contribution,  and all other cases with a singularity at 
$r=-1/(\alpha_{V1}^M +1/4)$ and a leading linear asymptotic behaviour.
A completely analogous discussion applies to the gravitational ghost term
$\cT^{\rm ghost}$ with the only difference that the type-II corresponds 
to $\alpha_{V}^G =+1/4$, and in a similar way to the first line in 
\eqref{gravTT} (type II corresponds to $\alpha_{T}^G =-1/6$).

Last but not least, in order to obtain a global solution the moving singularities
in the second line of \eqref{gravTT}  and in both expressions in  \eqref{gravsinv}
need to cancel against the numerators (Frobenius method). 
Even if this is arranged then these 
three terms have leading quadratic behaviour. However, there is one way 
to avoid this: if $\varphi^\star \propto r^2$ then 
$\varphi^{\star \, \prime\prime\prime}  \to 0$ for  $r\to \infty$, and 
the remaining two terms can be tuned to cancel.

To summarise this discussion, especially with respect to the impact of 
matter on the asymptotic behaviour, one notes that for type-II coarse-graining
the generic leading behaviour on the right hand side of the flow equation
is quadratic. Noting that the solution
of the differential equation $4 \varphi(r) -2 r \varphi '(r) = c r^2$ is 
$\varphi = \frac 1 2  c r^2\ln r  +{\cal O}(r^2) $
 this implies that the 
leading behaviour is then $\varphi^\star \propto r^2\ln r $ for $r\gg 1$.
As an advantage one has that then matter does not introduce any new 
singularities, {\it i.e.}, the same counting of conditions with respect to the
solubility and the number of solutions for this non-linear differential 
equation applies. For generic endomorphism parameters the leading asymptotic
behaviour of the matter contributions is linear, and thus will not qualitatively
change the leading asymptotic behaviour of the solution in the pure gravity
case. On the other hand, one introduces (even if one sets 
$\alpha_S^M=\alpha_{V2}^M$ right away) one or two new singularities
which will make without fine-tuning ({\it e.g.}, to push them to values of 
$r$ in which one is not interested, foremost to negative values) the differential
equation only locally solvable. Note that for the transverse vectors and for the 
Dirac fermions type-I endomorphism parameters behave for this 
purpose alike general values. 

For the scalars and the gauge ghosts the type-I and type-II endomorphism
parameters coincide, $\alpha_S^M=\alpha_{V2}^M=0$. Therefore, 
the ``dangers'' of type-II apply for the related  two terms also for 
type-I coarse-graining.  
At this point, it is 
interesting to note that, had one employed an interpolation scheme 
based on the Euler-MacLaurin formula,  the scalar term had simplified very much
alike the fermionic one does in the here used averaging interpolation
\cite{Alkofer:2018fxj,PhDthesis}:
\be
\cT^{\rm scalar} =  \frac {N_S} 2
 \frac 1 {(4\pi)^2}  (1+(\alpha_S^M+\tfrac 1 3 ) r ) \, .
\ee
With this behaviour the contributions of the scalars would be as 
easily and semi-analytically taken into account as the ones 
for the fermions here.

\subsection{Numerical solutions for global fixed functions}

\begin{figure*}[ht!]
\includegraphics[width=0.47\textwidth]{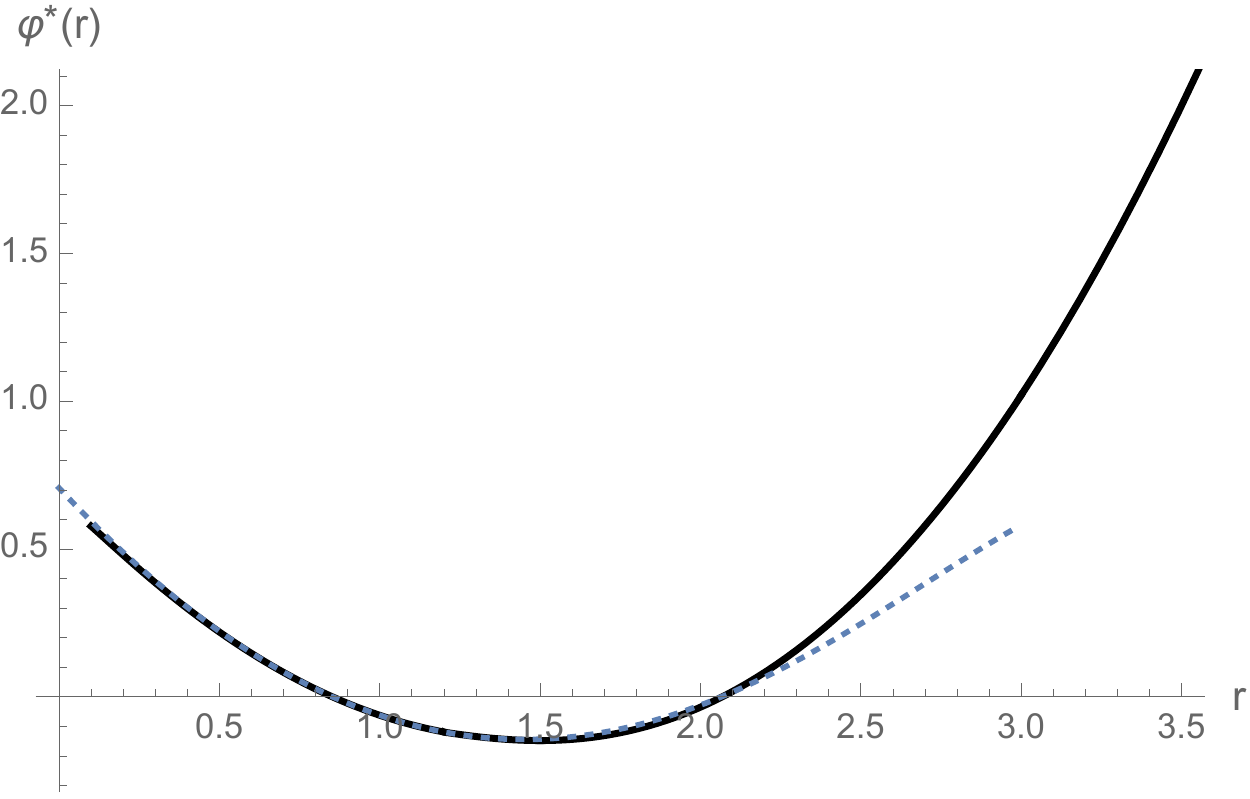} \hfill
\includegraphics[width=0.47\textwidth]{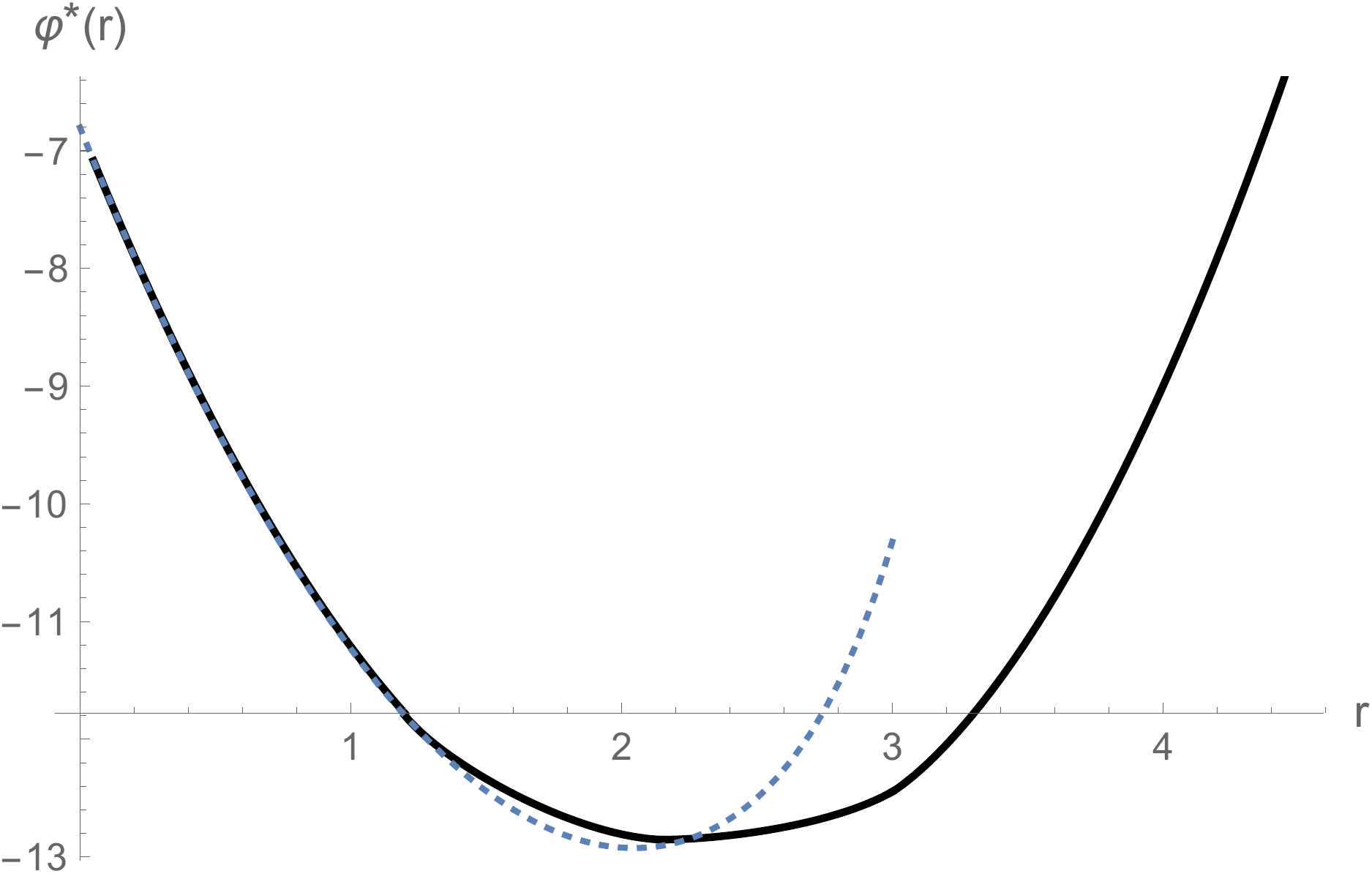}
\caption{\label{FigNumSol} 
Displayed  are the fixed functions (black lines) for 
the case of pure gravity (left panel) and with standard model matter
content (right panel). The respective  polynomial approximations of 
order 14  using $r=0$ as
an expansion point \cite{Alkofer:2018fxj} are shown  as dashed lines for comparison.}
\end{figure*}

In this section two examples for a numerical solution will be presented,
one for pure gravity and one for standard model matter content. 
Given the fact that type-II coarse graining with standard model matter content 
will lead to physically unacceptable solutions 
one may want to employ as coarse graining operator only the Laplacian 
(type-I). However, then already in
the pure gravity case the flow equation will not possess a solution for all 
positive curvatures $r$. 

The flow equation is a third order equation, and it is only then not over-constraint 
if there are a most three singularities \cite{Dietz:2012ic}.
Therefore, if the solution had no extremum, and there were no moving singularity 
one can allow for positive $r$ at most three fixed singularities. 
However, the physical condition of a positive Newton constant
and thus a negative $g_1$ implies that $\varphi(r)$ 
decreases at small values of the curvature.
On the other hand, at large curvatures the function $\varphi (r)$ should assume 
a positive value to make the functional integral well-defined which is achieved by 
$\varphi(r)\to + \infty$ for $r\to \infty$. Consequently, $\varphi^\star (r)$ 
must possess at least one minimum, and one can allow for at most two fixed 
singularities. However, for type-I one has four additional fixed singularities at
$r^{\rm sing}_2 =0$,  $r^{\rm sing}_3 =6/5$, $r^{\rm sing}_4 =3$ and  
$r^{\rm sing}_{\rm ghost} =4$,
where the last one originates from the ghost term $\cT^{\rm ghost}$.  
Searching for solutions for strictly positive curvature one does not require 
a condition at $r^{\rm sing}_2 =0$. The ghost singularity we move to negative
values of $r$ by choosing $\alpha_V^G=1/2$ which is well within the allowed 
range of parameters \cite{Ohta:2015fcu,Alkofer:2018fxj}. This is then 
the least modification of the flow equation as compared to the one 
in type-I coarse graining which allows for a numerical solution.

Analysing the flow equation for large $r$ one can infer the behaviour
\be
\label{AsympLog}
\varphi^\star (r)  \propto (2 \ln r - 1) \, r^2 + \cO \left( \frac {r^2}{\ln r} ,
\, \, r\, (\ln r)^2 \right)
\, ,
\ee
which fulfils the orders $r^2 (\ln r)^3$ and $r^2 (\ln r)^2$ simultaneously.
However, this asymptotic behaviour only becomes reasonably precise at 
extremely large values of $r$ and is thus only of limited use in the numerics.

To obtain numerical solutions  a multi-shooting method will be employed.
As for the pure gravity case:
to this end a minimum at the zero of the second summand in 
$\cT^{\rm TT}$ \eqref{gravTT} at $r=3/2$ is enforced. Shooting to the left 
one can then construct the solution left and right from the singularity by matching   
it at $r^{\rm sing}_3\pm 10^{-4} = 6/5 \pm 10^{-4}$ such that 
a singularity of the third derivative is avoided.
The result is displayed in the left panel of fig.\ \ref{FigNumSol}.

When adding the standard matter model content ($N_S=4, N_D=22.5$
and $N_V=12$ and type-I coarse graining) 
a polynomial approximation is used as an Ansatz in the differential
equation for the fixed function to estimate at which position the minimum
of $\varphi^\star(r)$ has to be located. This estimate is then iteratively
improved by repeating the
analogous procedure as in the gravity case. 

The result is displayed 
in the right panel of fig.~\ref{FigNumSol}.
First, one observes clearly the absence of any structure,
the global fixed functions are very close to parabolas, {\it i.e.},
all their features can be captured a quadratic expression with only 
three coefficients. Second, the agreement with the polynomial approximation
extends until $r \approx 2$. For the pure gravity case this implies that the
position of the minimum coincides with the one of the polynomial approximation
within numerical accuracy. For the matter-gravity system the minimum is
slightly shifted from $r=2.05$ to $r=2.15$.

\begin{figure*}[ht!]
\includegraphics[width=0.47\textwidth]{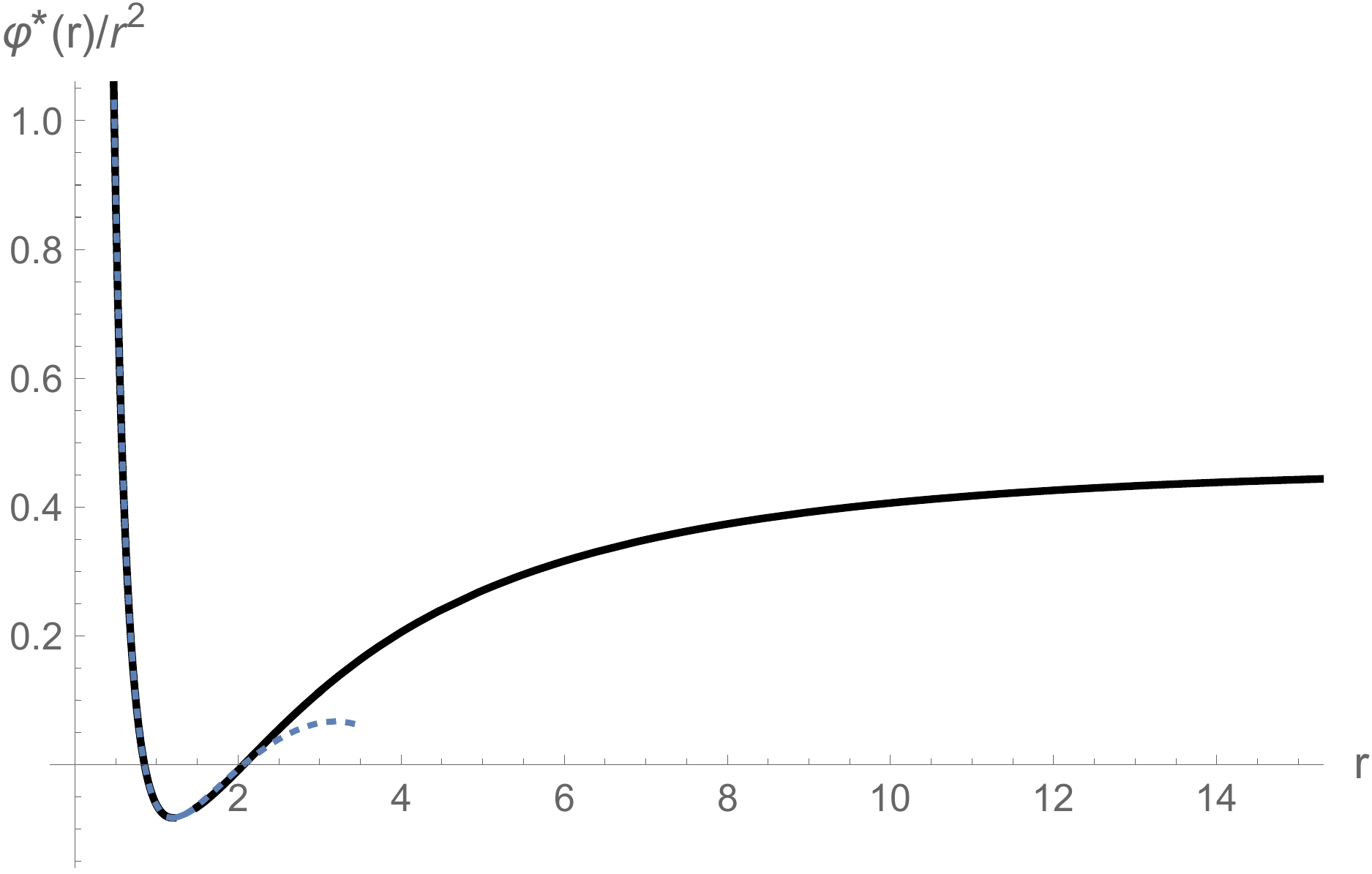} \hfill
\includegraphics[width=0.47\textwidth]{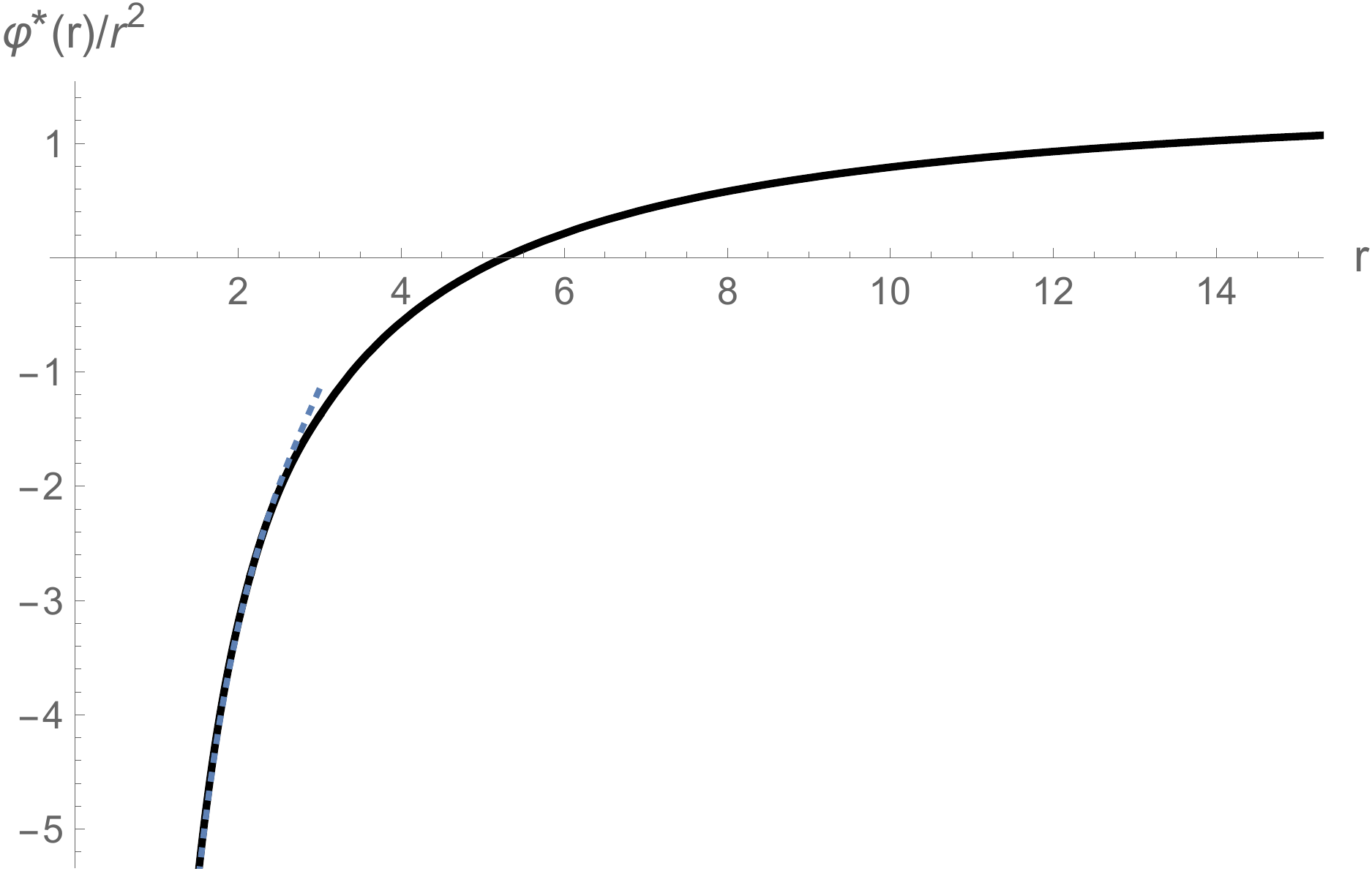}
\caption{\label{FigNumSol2} 
Displayed  are the effective potentials $\varphi^\star (r) / r^2 $ (black lines) for 
the case of pure gravity (left panel) and with standard model matter
content (right panel). The respective  polynomial approximations of 
order 14  using $r=0$ as
an expansion point \cite{Alkofer:2018fxj} are shown  as dashed lines for comparison.}
\end{figure*}

In fig.\ \ref{FigNumSol2} the functions $\varphi^\star (r) / r^2 $ are plotted
for two reasons. First, in this way one can check the asymptotic behaviour for 
large $r$. Amazingly, the logarithmic dependence of eq.\ \eqref{AsympLog} cannot 
be seen. Even up to very large values of $r$ the extracted 
leading term is proportional to $r^2$ within the numerical accuracy. Second, 
plotting the fixed functions in this way a comparison to the results of 
ref.\ \cite{Christiansen:2017bsy} is straightforwardly
possible. In this investigation fixed functions with 
a sphere as  background have been calculated, however, using a linear split of 
the metric and a vertex expansion. As argued in this reference the ratio 
$\varphi^\star (r) / r^2 $ is the effective background potential at the 
fixed point, and a minimum of this function signals a solution of the
background equation of motion. The authors of ref.\ \cite{Christiansen:2017bsy}
found now in the pure gravity case a background potential without an extremum 
whereas after adding the Standard Model matter content they observed a minimum
at $r \approx 0.1$. From the present work, see fig.\ \ref{FigNumSol2}, one 
obtains, on the contrary, a minimum for the pure gravity case
and the absence of an extremum with Standard Model matter. In addition, 
in the pure gravity case the minimum is at a rather large value of the 
curvature, $r\approx 1.2$.

\section{Summary and Conclusions}
\label{sec:Concl}

In this letter a study of global fixed functions
in the context of the asymptotic  safety scenario for $f(R)$-gravity 
coupled to matter has been presented. 
For some well-chosen sets of coarse-graining 
parameters global quadratic solution exists.  Although these 
choices can hardly be motivated by physics 
they explain a remarkable behaviour of the 
numerically obtained global fixed functions: for all studied 
cases the deviations from a global quadratic form are 
tiny. Given this situation one might even speculate that differences
to a quite simple form of the global fixed functions might be 
only due to the employed truncation.  
In the present investigation only two relevant directions have been found, 
one of them is directly related to the constant term,
{\it i.e.}, the cosmological constant. The other one is with only very
small contributions from higher-order terms a linear
combination of a linear and a quadratic term. 

The presented investigation emphasises once more the question whether  
a change of a 
coarse-graining operator by a non-trivial endomorphism parameter still 
leads to consider the same theory, or whether the ultraviolet completions
of such quantum gravity models are qualitatively different. 
Searching for global solutions for the fixed functions
provided further evidence for the conclusion of 
ref.\ \cite{Alkofer:2018fxj} based on polynomial approximations,
namely, that the NGFPs seen in gravity-matter systems belong to 
(at least) two different classes, and that the use of these two different 
schemes lead to different quantisation prescriptions for the same
``classical'' theory. Such a situation deserves certainly further 
investigations.

When comparing the here obtained numerical solutions for the 
fixed functions with those of ref.\ \cite{Christiansen:2017bsy} 
a clear difference can be noted. The vertex expansion used there
is certainly a more sophisticated truncation of the flow equation
than the single-metric background approach used in this work.
On the other hand, the exponential split of the metric employed here
might be from a conceptual point of view superior to the 
linear split, {\it cf.}, also the recent study \cite{deBrito:2018jxt}
where (using also the same gauge as in the presented calculation) 
a two-parameter family of parameterisations of the split of the
metric  has been applied to the pure gravity case. Depending
on these parameters two different classes for the fixed functions 
have been found. One might now further investigate whether such
differences persist when matter is included, and how the assignment 
of these different classes relate to the different classes found 
 in ref.\  \cite{Alkofer:2018fxj} and here.

With respect to an investigation of the truncation dependence 
of the NGFP functions one may build on the recent work of ref.\ 
\cite{Lippoldt:2018wvi}: there a flow equation has been
constructed retaining the consistency of the fluctuation field and 
background field equations of motion even for finite 
RG scales. Within the background approximation this leads to a
modified flow equation containing some additional terms. It 
is certainly worthwhile to study whether including these terms 
brings the results for the fixed functions closer to the ones 
found in a vertex expansion for background and fluctuating 
fields.

Last but not least, the results for the fixed functions obtained here 
verify further what one 
has seen in practically all investigations of the NGFP function for 
$f(R)$-gravity without and with matter: if a solution for such a function 
can be found it is very close to a polynomial of only quadratic order.



\section*{Acknowledgments}
I thank Frank Saueressig for the collaboration on the project published in ref.\ 
\cite{Alkofer:2018fxj} which provided the foundation for the work presented here.
Special thanks go to
Jan M.\ Pawlowski and  Frank Saueressig  for their critical reading of this manuscript
and helpful comments.  
Furthermore, I am grateful to Holger Gies, Benjamin Knorr, Daniel F.\ Litim, 
Jan M.\ Pawlowski, Roberto Percacci and Antonio D.\ Pereira for inspiring discussions. 
\\
I acknowledge financial support by the Netherlands Organization for Scientific
Research (NWO) within the Foundation for Fundamental Research on Matter (FOM) grants
13PR3137 and 13VP12. 


\begin{thebibliography}{10}

\bibitem{Reuter:1996cp}
  M.~Reuter,
  Phys.\ Rev.\ D {\bf 57} (1998) 971
  [hep-th/9605030].

\bibitem{Alkofer:2018fxj}
  N.~Alkofer and F.~Saueressig,
  Annals Phys.\  {\bf 396} (2018) 173
  [arXiv:1802.00498].

\bibitem{PhDthesis}
N.~Alkofer, 
PhD thesis, University of Nijmegen, 2018, ISBN 978-94-92896-57-5.

\bibitem{Ohta:2015efa}
  N.~Ohta, R.~Percacci and G.~P.~Vacca,
  Phys.\ Rev.\ D {\bf 92} (2015) 
  061501
  [arXiv:1507.00968].

\bibitem{Ohta:2015fcu} 
  N.~Ohta, R.~Percacci and G.~P.~Vacca,
  Eur.\ Phys.\ J.\ C {\bf 76} 
(2016) 46 
  [arXiv:1511.09393].

\bibitem{Reuter:2008qx}
M.~Reuter and H.~Weyer,
Phys.\ Rev.\ D {\bf 80} (2009) 025001 
[arXiv:0804.1475].

\bibitem{Benedetti:2012dx}
D.~Benedetti and F.~Caravelli,
JHEP {\bf 06} (2012) 017
Erratum: [JHEP {\bf 10} (2012) 157] 
[arXiv:1204.3541].

\bibitem{Demmel:2012ub}
  M.~Demmel, F.~Saueressig and O.~Zanusso,
  JHEP {\bf 11} (2012) 131 
  [arXiv:1208.2038].

\bibitem{Dietz:2012ic}
J.~A.~Dietz and T.~R.~Morris,
JHEP {\bf 01} (2013) 108 
[arXiv:1211.0955].

\bibitem{Bridle:2013sra}
I.~H.~Bridle, J.~A.~Dietz and T.~R.~Morris,
JHEP {\bf 03} (2014) 093 
[arXiv:1312.2846].

\bibitem{Dietz:2013sba}
J.~A.~Dietz and T.~R.~Morris,
JHEP {\bf 07} (2013) 064 
[arXiv:1306.1223].


\bibitem{Demmel:2014sga}
  M.~Demmel, F.~Saueressig and O.~Zanusso,
  JHEP {\bf 06} (2014) 026 
  [arXiv:1401.5495].

\bibitem{Demmel:2014hla}
  M.~Demmel, F.~Saueressig and O.~Zanusso,
  Annals Phys.\  {\bf 359} (2015) 141 
  [arXiv:1412.7207].

\bibitem{Demmel:2015oqa}
  M.~Demmel, F.~Saueressig and O.~Zanusso,
  JHEP {\bf 08} (2015) 113 
  [arXiv:1504.07656].

\bibitem{Labus:2016lkh}
P.~Labus, T.~R.~Morris and Z.~H.~Slade,
Phys.\ Rev.\ D {\bf 94} (2016) 024007 
[arXiv:1603.04772].

\bibitem{Dietz:2016gzg}
J.~A.~Dietz, T.~R.~Morris and Z.~H.~Slade,
Phys.\ Rev.\ D {\bf 94} (2016) 124014 
[arXiv:1605.07636].


\bibitem{Knorr:2017mhu}
  B.~Knorr,
  Class.\ Quant.\ Grav.\  {\bf 35} (2018) 
115005
  [arXiv:1710.07055].

\bibitem{Christiansen:2017bsy}
  N.~Christiansen, K.~Falls, J.~M.~Pawlowski and M.~Reichert,
  Phys.\ Rev.\ D {\bf 97} (2018) 
046007
  [arXiv:1711.09259].

\bibitem{Falls:2017lst}
  K.~Falls, C.~R.~King, D.~F.~Litim, K.~Nikolakopoulos and C.~Rahmede,
  Phys.\ Rev.\ D {\bf 97} (2018) 
086006
  [arXiv:1801.00162].


\bibitem{deBrito:2018jxt}
  G.~P.~De Brito, N.~Ohta, A.~D.~Pereira, A.~A.~Tomaz and M.~Yamada,
  Phys.\ Rev.\ D {\bf 98} (2018) 
026027
  [arXiv:1805.09656].

\bibitem{Dou:1997fg}
  D.~Dou and R.~Percacci,
  Class.\ Quant.\ Grav.\  {\bf 15} (1998) 3449
  [hep-th/9707239].

\bibitem{Percacci:2002ie}
  R.~Percacci and D.~Perini,
  Phys.\ Rev.\ D {\bf 67} (2003) 081503
  [hep-th/0207033].

\bibitem{Narain:2009fy}
  G.~Narain and R.~Percacci,
  Class.\ Quant.\ Grav.\  {\bf 27} (2010) 075001 
  [arXiv:0911.0386].

\bibitem{Daum:2010bc}
  J.-E.~Daum, U.~Harst and M.~Reuter,
  Gen.\ Rel.\ Grav.\  {\bf 43} (2011) 2393 
  [arXiv:1005.1488].

\bibitem{Folkerts:2011jz}
  S.~Folkerts, D.~F.~Litim and J.~M.~Pawlowski,
  Phys.\ Lett.\ B {\bf 709} (2012) 234 
  [arXiv:1101.5552].

\bibitem{Harst:2011zx}
  U.~Harst and M.~Reuter,
  JHEP {\bf 05} (2011) 119 
  [arXiv:1101.6007].

\bibitem{Eichhorn:2011pc}
  A.~Eichhorn and H.~Gies,
  New J.\ Phys.\  {\bf 13} (2011) 125012 
  [arXiv:1104.5366].

\bibitem{Eichhorn:2012va}
  A.~Eichhorn,
  Phys.\ Rev.\ D {\bf 86} (2012) 105021 
  [arXiv:1204.0965].

\bibitem{Dona:2012am}
  P.~Don{\`a} and R.~Percacci,
  Phys.\ Rev.\ D {\bf 87} (2013) 045002 
  [arXiv:1209.3649].

\bibitem{Henz:2013oxa}
  T.~Henz, J.~M.~Pawlowski, A.~Rodigast and C.~Wetterich,
  Phys.\ Lett.\ B {\bf 727} (2013) 298 
  [arXiv:1304.7743].

\bibitem{Dona:2013qba}
  P.~Don{\`a}, A.~Eichhorn and R.~Percacci,
  Phys.\ Rev.\ D {\bf 89} (2014) 084035 
  [arXiv:1311.2898].

\bibitem{Percacci:2015wwa}
  R.~Percacci and G.~P.~Vacca,
  Eur.\ Phys.\ J.\ C {\bf 75} (2015) 188 
  [arXiv:1501.00888].

\bibitem{Labus:2015ska}
  P.~Labus, R.~Percacci and G.~P.~Vacca,
  Phys.\ Lett.\ B {\bf 753} (2016) 274 
  [arXiv:1505.05393].

\bibitem{Oda:2015sma}
  K.~y.~Oda and M.~Yamada,
  Class.\ Quant.\ Grav.\  {\bf 33} (2016) 125011 
  [arXiv:1510.03734].

\bibitem{Meibohm:2015twa}
  J.~Meibohm, J.~M.~Pawlowski and M.~Reichert,
  Phys.\ Rev.\ D {\bf 93} (2016) 084035 
  [arXiv:1510.07018].


\bibitem{Dona:2015tnf}
  P.~Don{\`a}, A.~Eichhorn, P.~Labus and R.~Percacci,
  Phys.\ Rev.\ D {\bf 93} (2016) 044049, 
   Erratum: [Phys.\ Rev.\ D {\bf 93} (2016) 
129904],
  [arXiv:1512.01589].

\bibitem{Meibohm:2016mkp}
  J.~Meibohm and J.~M.~Pawlowski,
  Eur.\ Phys.\ J.\ C {\bf 76} (2016) 285 
  [arXiv:1601.04597].

\bibitem{Eichhorn:2016esv}
  A.~Eichhorn, A.~Held, J.~M.~Pawlowski,
  Phys.\ Rev.\ D {\bf 94} (2016) 104027
  [arXiv:1604.02041].

\bibitem{Henz:2016aoh}
  T.~Henz, J.~M.~Pawlowski and C.~Wetterich,
  Phys.\ Lett.\ B {\bf 769} (2017) 105 
  [arXiv:1605.01858].

\bibitem{Eichhorn:2016vvy}
  A.~Eichhorn and S.~Lippoldt,
  Phys.\ Lett.\ B {\bf 767} (2017) 142 
  [arXiv:1611.05878].

\bibitem{Christiansen:2017gtg}
  N.~Christiansen and A.~Eichhorn,
  Phys.\ Lett.\ B {\bf 770} (2017) 154 
  [arXiv:1702.07724].

\bibitem{Eichhorn:2017eht}
  A.~Eichhorn and A.~Held,
  Phys.\ Rev.\ D {\bf 96} (2017) 086025 
  [arXiv:1705.02342].

\bibitem{Christiansen:2017qca}
  N.~Christiansen, A.~Eichhorn and A.~Held,
  Phys.\ Rev.\ D {\bf 96} (2017) 084021 
  [arXiv:1705.01858].

\bibitem{Eichhorn:2017ylw}
  A.~Eichhorn and A.~Held,
  Phys.\ Lett.\ B {\bf 777} (2018) 217 
  [arXiv:1707.01107].

\bibitem{Eichhorn:2017lry}
  A.~Eichhorn and F.~Versteegen,
  JHEP {\bf 01} (2018) 030 
  [arXiv:1709.07252].

\bibitem{Eichhorn:2017egq}
  A.~Eichhorn,
  arXiv:1709.03696.

\bibitem{Christiansen:2017cxa}
  N.~Christiansen, D.~F.~Litim, J.~M.~Pawlowski and M.~Reichert,
  Phys.\ Rev.\ D {\bf 97} (2018) 
106012
  [arXiv:1710.04669].

\bibitem{Wetterich:1992yh}
  C.~Wetterich,
  Phys.\ Lett.\ B {\bf 301} (1993) 90 
  [arXiv:1710.05815].

\bibitem{Morris:1993qb}
  T.~R.~Morris,
  Int.\ J.\ Mod.\ Phys.\ A {\bf 9} (1994) 2411 
  [hep-ph/9308265].


\bibitem{Rubin:1984tc}
  M.~A.~Rubin and C.~R.~Ordonez,
  J.\ Math.\ Phys.\  {\bf 25} (1984) 2888;
  J.\ Math.\ Phys.\  {\bf 26} (1985) 65.

\bibitem{Litim:2001up}
  D.~F.~Litim,
  Phys.\ Rev.\ D {\bf 64} (2001) 105007
  [hep-th/0103195].

\bibitem{Litim:2000ci} 
  D.~F.~Litim,
  Phys.\ Lett.\ B {\bf 486} (2000) 92
  [hep-th/0005245].

\bibitem{Codello:2008vh}
  A.~Codello, R.~Percacci and C.~Rahmede,
  Annals Phys.\  {\bf 324} (2009) 414 
  [arXiv:0805.2909 [hep-th]].

\bibitem{Lippoldt:2018wvi}
  S.~Lippoldt,
  Phys.\ Lett.\ B {\bf 782} (2018) 275
  [arXiv:1804.04409 [hep-th]].


\end{thebibliography}
\end{document}